\documentclass[aps,prl,twocolumn,showpacs,amsmath,amssymb,superscriptaddress]{revtex4}
\usepackage{graphicx}
\usepackage{latexsym}
\usepackage{amsmath}
\usepackage{color}
\usepackage{graphicx}% Include figure files
\usepackage{dcolumn}% Align table columns on decimal point
\usepackage{bm}
\usepackage{mathrsfs}
\usepackage{feynmf}
\usepackage{comment}
\usepackage{float}

\def\d{\textrm{d}}
\def\cos{\textrm{cos}}

\newcommand{\beq}{\begin{equation}}
\newcommand{\eeq}{\end{equation}}
\newcommand{\ba}{\begin{array}{ccc}}
\newcommand{\ea}{\end{array}}
\newcommand{\nn}{\nonumber \\}

\newcommand{\br}{{\bm r}}

\newcommand{\bk}{{\bm k}}
\newcommand{\bp}{{\bm p}}

\newcommand{\bq}{{\bm q}}

\def\bea{\begin{eqnarray}}
\def\eea{\end{eqnarray}}

\begin{document}

\title{Wess-Zumino-Witten Terms in Graphene Landau Levels}

\author{Junhyun Lee}
%\email{junhyunlee@fas.harvard.edu}
\affiliation{Department of Physics, Harvard University, Cambridge MA 02138, USA}

\author{Subir Sachdev}
%\email{sachdev@physics.harvard.edu}
\affiliation{Department of Physics, Harvard University, Cambridge MA 02138, USA}
\affiliation{Perimeter Institute for Theoretical Physics, Waterloo, Ontario N2L 2Y5, Canada}

\date{\today}

\begin{abstract}
We consider the interplay between the antiferromagnetic and Kekul\'e valence bond solid orderings
in the zero energy Landau levels of neutral monolayer and bilayer graphene. We establish the presence of Wess-Zumino-Witten
terms between these orders: this implies that their quantum fluctuations are described by the deconfined critical theories
of quantum spin systems. We present implications for experiments, including the possible presence of excitonic superfluidity
in bilayer graphene.
\end{abstract}


\maketitle

%\section{Introduction}

\section*{Introduction}

\begin{figure}
\includegraphics[width=3.6in]{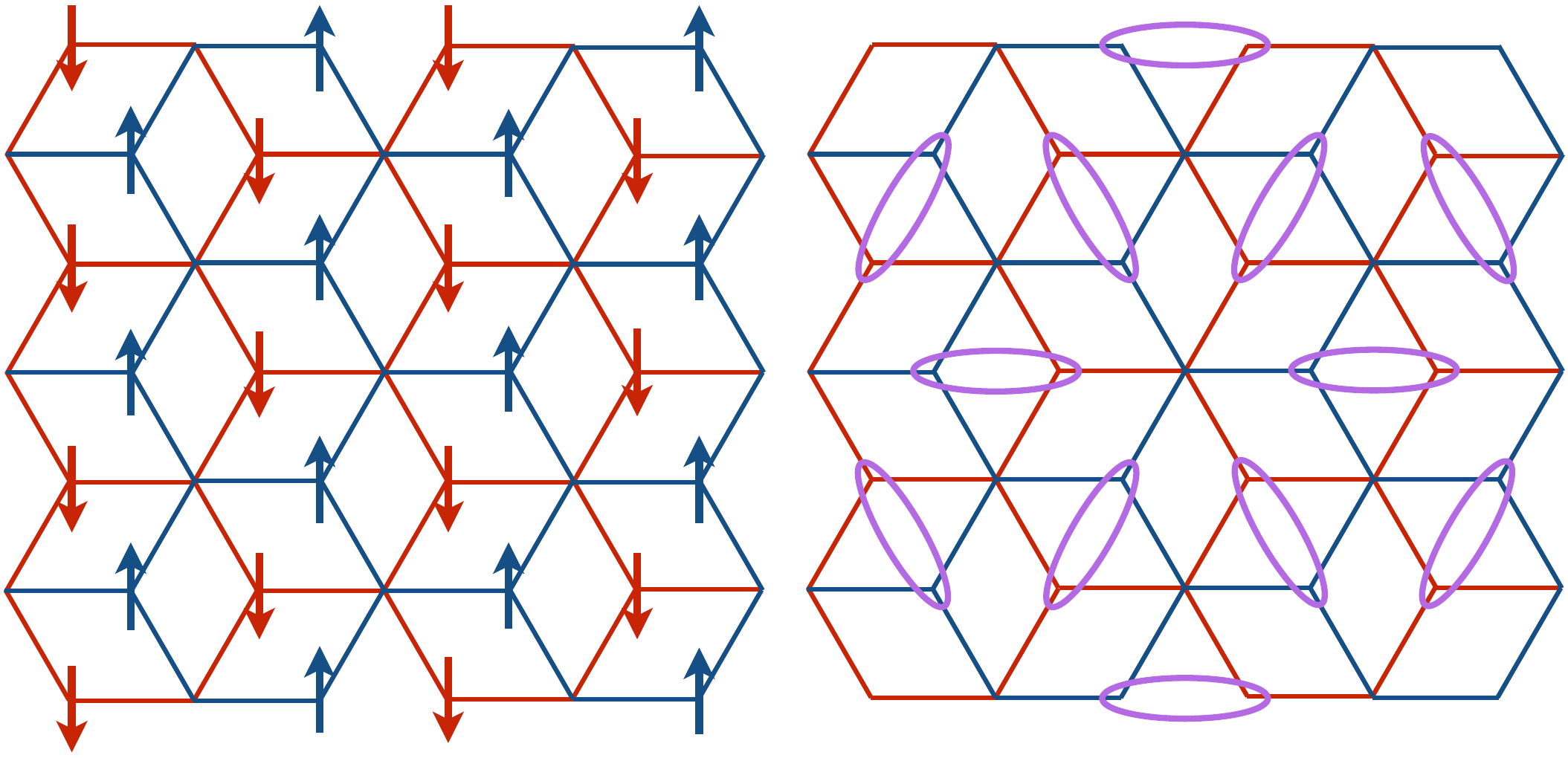}
\caption{AF (left) and Kekul\'e VBS states of bilayer graphene. The blue (red) lines indicate the honeycomb lattice of the top (bottom) layer.
The ellipses in the VBS state denote the links between the top and bottom layers which are equivalently distorted with respect to the parent lattice.}
\label{fig:blg}
\end{figure}

A number of recent experimental \cite{weitz,freitag,macdonaldexp,young1,maher,basel,young2} and 
theoretical \cite{herbut1,jung,vafek,allan,levitov,khari1,khariprl,khari2,khari3,so5,JLSS14,efrat} works have focused on the
presence of antiferromagnetism in neutral monolayer and bilayer graphene
in an applied magnetic field. It has also been argued that a nonmagnetic state with lattice symmetry breaking
in the Kekul\'e valence bond solid (VBS) pattern (see Fig.~\ref{fig:blg}) is proximate to the antiferromagnetic (AF) state \cite{khari1,khariprl,so5,JLSS14}.
Bilayer graphene offers a particularly attractive area for studying the interplay between the AF and VBS
order because it may be possible to tune between them by applying a transverse electric field \cite{weitz,khariprl,JLSS14}.

The presence of the competing AF and VBS orders sets up the possibility \cite{JLSS14} of novel quantum criticality
between these orders, similar to that found in insulating quantum spin models \cite{vbsprb,senthil1,senthil2,clark,lauchli,ganesh,zhu,block,shengfisher,damle,langsun}. However these quantum spin models apply in the limit of very large
on-site Coulomb repulsion between the electrons, and this is not the appropriate parameter regime for graphene.
Here we examine a complementary limit of large magnetic field and moderate interactions, so that it is permissible to project
onto an effective Hamiltonian acting only on the zero energy Landau levels. Such a limit has been widely
used with considerable success in describing the properties of graphene. (Note, however, that we are still in the regime where the cyclotron gap is still smaller than the tight-binding hopping parameters, with magnetic fields smaller than 10 T.) Our main new result is that the Landau level projected effective
action for the AF and VBS orders has a topological Wess-Zumino-Witten (WZW) term \cite{WZ,Witten,abanov} for both the monolayer
and bilayer cases. 

The WZW term has a quantized coefficient, and it
computes a Berry phase linking together spatial and temporal textures in the AF and VBS orders. It can be viewed as a higher dimensional
generalization of the Berry phase of a single spin $S$ degree of freedom, which is equal to $S$ times the area enclosed by the spin worldline
on the unit sphere. Similarly, the WZW term here measures the area on the surface of the sphere in the five-dimensional AF and VBS order parameter space.
The presence of this term implies \cite{tanakahu,senthil3,grover2} that the field theories of deconfined criticality \cite{senthil1,senthil2} 
apply to graphene. Such theories describe the quantum phase transition not in the conventional Landau terms of fluctuating order parameters,
but using fractionalized degrees of freedom coupled to emergent gauge fields.
We will also discuss experimental implications of these results

\section*{Model and results}

We begin by directly stating the Hamiltonian of the low energy graphene bands  (see {\em e.g.\/} Refs.~\cite{khari3,JLSS14} for details)
\beq
H = v \left( \begin{array}{cc} 0 & a^q \\ a^{\dagger q} & 0 \end{array} \right)
\eeq
where $v$ is a Fermi velocity, $a= p_x - i p_y - (e/c)(A_x - i A_y)$ with $(p_x,p_y)$ the electron momentum, $(A_x, A_y)$ is the vector potential of the applied
magnetic field, the matrix acts on the graphene sublattice index, and $q=1$ for monolayer graphene, while $q=2$ for the bilayer case.
For bilayers, the sublattice index coincides with the layer index. For both monolayers and bilayers, there is an additional twofold valley degeneracy,
along with the usual twofold spin degeneracy (in the absence of a Zeeman coupling). 

The $a$, $a^\dagger$ obey commutation relations proportional to those of the ladder operators of a harmonic oscillator, and so
it is easy to diagonalize $H$. In this manner we obtain $q$ zero energy Landau levels, which are spanned by the orthonormal 
eigenfunctions $\psi_\ell (\br)$, where $\ell = 1, \ldots, q N_\Phi$, with $N_\Phi$ the number of flux quanta. So we write the electron
annihilation field operator projected to the zero energy Landau levels as
\beq
\Psi (\br) = \sum_{\ell=1}^{q N_\Phi} \psi_\ell (\br) c_\ell ,
\label{Psic}
\eeq
where $c_\ell$ is canonical fermion annihilation operator. 
In the zero energy Landau levels, the valley, sublattice, and layer indices all coincide; henceforth we will refer to this as a valley index, and it
can take two values. The fermion operators also carry a spin index with two possible values, and we do not explicitly display the spin or valley indices.

We now introduce Pauli matrices $\sigma^{x,y,z}$ which acts on the spin space,
and a second set $\rho^{x,y,z}$ which act on the valley space (here we follow the conventions of Ref.~\cite{JLSS14}).
In terms of these matrices, the three-component AF order is measured by $(\rho^z \sigma^x, \rho^z \sigma^y, \rho^z \sigma^z)$
while the two-component VBS order parameter is $(\rho^x, \rho^y)$.

It is convenient to write the above matrices as
\beq
\Gamma_1 = \rho^z \sigma^x ~,~\Gamma_2 = \rho^z \sigma^y~,~\Gamma_3 = \rho^z \sigma^z ~,~\Gamma_4 = \rho^x
~,~\Gamma_5 = \rho^y \nonumber
\eeq
and to notice that the 5 $\Gamma_a$ matrices anticommute and square to unity; indeed these are the 5 Dirac gamma matrices. 
Their 10 products $i \Gamma_a \Gamma_b$ ($a \neq b$) realize the Lie algebra of SO(5), 
and the 15 matrices $\Gamma_a$ and $i \Gamma_a \Gamma_b$ realize the Lie algebra of SU(4).

Next, we introduce a five-component unit vector $n_a (\br ,\tau)$, where $\br = (x,y)$ are the spatial coordinates and $\tau$ is imaginary time, 
representing the combined spacetime fluctuations of the AF
and VBS orders. Then the imaginary time Lagrangian of the electrons projected to the zero energy Landau levels is
\beq
\mathcal{L} = \sum_{\ell=1}^{q N_\Phi} c_\ell^\dagger  \frac{\partial c_\ell }{\partial \tau}
 - \lambda \int d^2 \br \,  n_a (\br, \tau) \, \Psi^\dagger (\br, \tau) \Gamma_a \Psi (\br, \tau)  \,,
 \label{L}
\eeq
where $\lambda$ is the coupling of the electrons to the AF and VBS orders,
and there is an implicit sum of $a$ over five values, and also over the spin and valley indices. The $\lambda$ term arises from
a decoupling of the electron-electron interactions specified in Refs.~\onlinecite{khari1,so5,JLSS14}.

Now we can state our primary result. We integrate over the the $c_\ell$ electrons in $\mathcal{L}$ and obtain an effective action
for unit vector $n_a (\br, \tau)$. Apart from the usual terms of the O(5) nonlinear sigma model considered in Ref.~\cite{so5} (and anisotropies due to the Zeeman coupling, electron-electron interactions, and a possible transverse electric field for bilayers), the effective action has a topological WZW term at level $q$,
\bea
\mathcal{S}_{\mathrm{WZW}} &=& 2 \pi i q \, W[n_a]  \label{wzw} \\
W [n_a] \! &=&\! \frac{3}{8\pi^2} \int_0^1 \! du \int \! \d^2 \br d \tau \epsilon_{abcde} n_a \partial_x n_b \partial_y n_c \partial_\tau n_d \partial_u n_e
\nonumber
\eea
Here we have introduced the extra coordinate $u$, and $n_a (\br, \tau, u)$ is any function which smoothly extrapolates from 
the physical $n_a (\br , \tau)$ at $u=1$ to a fixed value (say) $n_a = (1,0,0,0,0)$ at $u=0$. The choice of the extrapolation can only
change $W[n_a]$ by integers, and so $e^{2 \pi i q W}$ is well defined.

In the case of graphene in zero magnetic field and weak interactions, 
the same WZW term between the N\'eel and VBS orders is also present \cite{liang}. 
However, for the experimentally important case of bilayer graphene, there is no such WZW term for the AF and VBS orders at zero field and 
weak interactions \cite{JLSS14,egmoon} (although, E.-G.~Moon has noted such a term for the quantum spin Hall order \cite{egmoon});
so, in this case the zero energy Landau level projection is crucial for obtaining the topological coupling.

Such a WZW term has a strong impact in the interplay between the order parameters. As we will review below, it topologically links AF order to
defects of the VBS order, and vice versa.

\section*{Derivation}

We provide two derivations of Eq.~(\ref{wzw}). 

First, pick any three of the five $n_a$ components, say $a=u,v,w$, and set the other two to zero.
Then we have unit 3-vector field $\vec{N} = (n_u, n_v, n_w)$. Now consider a static Skyrmion texture in $\vec{N} (\br)$. 
Then by a computation parallel to that in Section III.B of K.~Moon {\em et al.} \cite{moon} (and its generalization to $q=2$ \cite{abanin}),
the Skyrmion acquires a ``charge.'' In the present situation the charge is measured by $i \Gamma_u \Gamma_v \Gamma_w$
and its spatial density is \cite{ref}
\beq
\left \langle \Psi^\dagger (\br) \,  i \Gamma_u \Gamma_v \Gamma_w \Psi (\br) \right \rangle  =  \frac{q}{2\pi} \vec{N} \cdot (\partial_x \vec{N} \times \partial_y \vec{N} )
\label{charge}
\eeq
where the angular brackets represent the expectation value over the occupied states in the zero energy Landau level perturbed 
by the texture in $\vec{N}$ as in $\mathcal{L}$. 
(A similar relationship has been noted in monolayer graphene in zero magnetic field \cite{herbut2}; however, no such relationship
applies to bilayer graphene in zero field.)
Now consider a VBS vortex, i.e., a $2 \pi$ vortex in $(n_4, n_5)$ applied to $\mathcal{L}$.
For a two-component order, the core of the vortex has a singularity, but this can be relieved by orienting $n_a$ in a third direction, say
$(\pm 1,0,0,0,0)$. Now the VBS vortex is equivalent to a half-Skyrmion in $\vec{N} = (n_1, n_4, n_5)$, and after integrating Eq.~(\ref{charge})
over all space, this vortex has $\langle \sigma^x \rangle = \pm q$. Similarly, vortex cores in the directions $(0,\pm 1, 0,0,0)$ and $(0,0,\pm 1, 0,0)$
yield $\langle \sigma^y \rangle = \pm q$ and $\langle \sigma^z \rangle = \pm q$. So we reach the important conclusion
that the VBS vortex has total spin $S=q/2$, and has an associated $(q+1)$-fold degeneracy. For $q=1$, note that this is precisely the situation
considered in Ref.~\cite{levin} for quantum spin models (see also Ref.~\cite{hou-prb}). Alternatively, we can examine the fate of $\mathcal{S}_{\mathrm{WZW}}$ in the presence of such VBS vortices: following a computation by Grover and Senthil \cite{grover2}, we find that the WZW term reduces to the
quantum Berry phase of a single spin with $S=q/2$. From this we conclude that Eq.~(\ref{charge}) implies Eq.~(\ref{wzw}).

For a second derivation of the WZW term from Eq.~(\ref{L}), we examine a diagrammatic expansion of $\mathcal{L}$. 
Consider a situation where $n_a$ is polarized near, say, $(0,0,0,0,1)$. Then, we can write $n_a = (\pi_1, \pi_2, \pi_3, \pi_4,1)$
where $|\pi_i| \ll 1$ for $i=1, \ldots, 4$. Then to zeroth order in the $\pi_i$, the $c_\ell$ operators in $\mathcal{L}$ have the Green's function
\beq
G = (i \omega + \lambda \Gamma_5)^{-1} \label{G}
\eeq
where $\omega$ is the frequency of the electron propagator. We now proceed to integrate out the electrons, and derive an effective
action for the $\pi_i$. 
\begin{figure}
\includegraphics[width=2.2in]{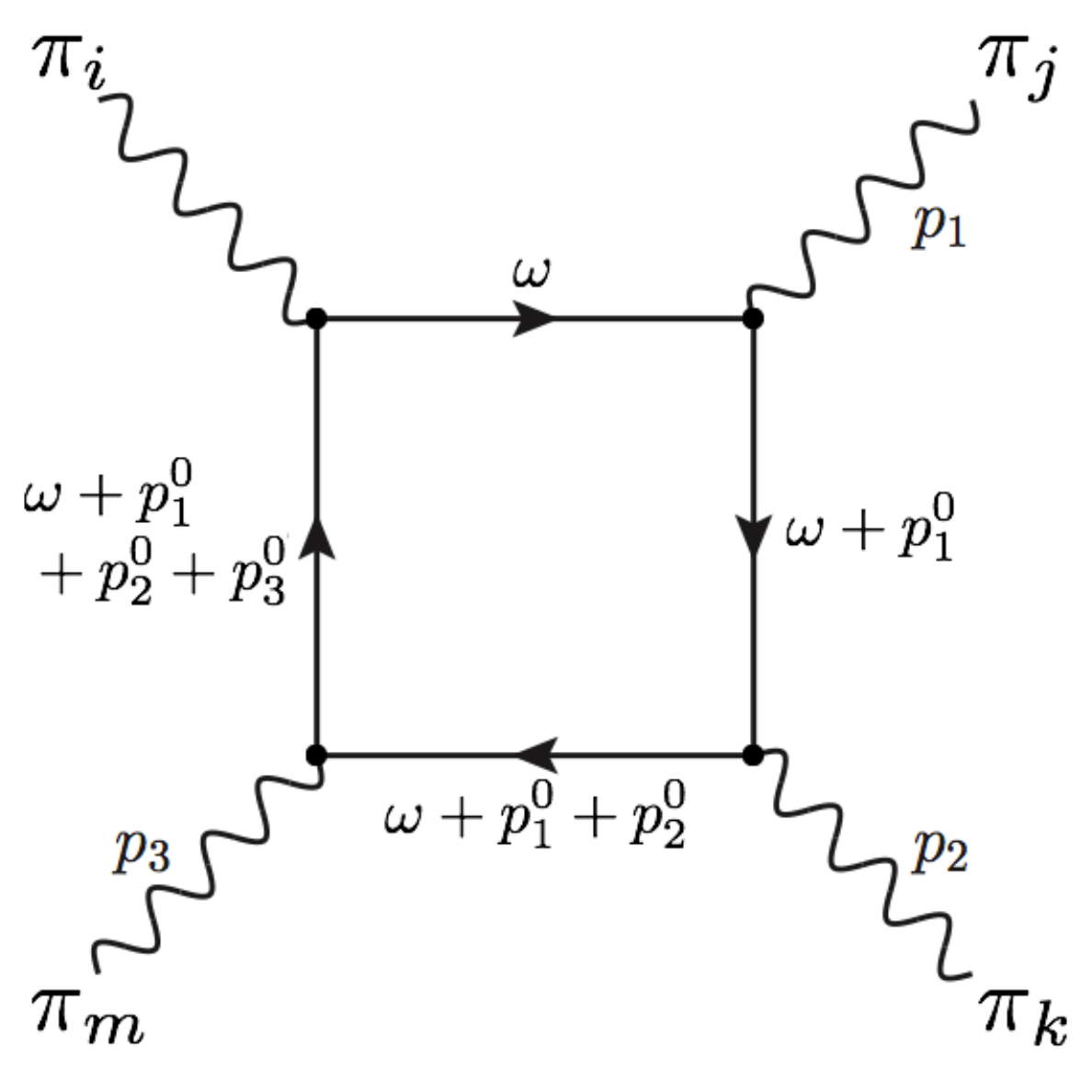}
\caption{Box diagram leading to $\mathcal{S}_\pi$. The full lines are the Green's function in Eq.~(\ref{G}) at the labeled 
frequencies, and the vertices are the $\lambda$ term in Eq.~(\ref{L}). }
\label{fig:box}
\end{figure}
At fourth order in the $\pi_i$, we consider the box diagram in Fig.~\ref{fig:box}; this can be evaluated by methods
similar to those in Ref.~\cite{JLSS14}, but with the $G$ above, and the vertices contributing the factors implied by Eq.~(\ref{L}). A computation described
in the Supplemental Material \cite{ref} yields the contribution
\beq
\mathcal{S}_\pi = \frac{i 3 q}{16 \pi} \int d^2 \br d \tau \epsilon_{ijkm} \pi_i \partial_x \pi_j \partial_y \pi_k \partial_\tau \pi_m \, .
\label{Spi}
\eeq
It can be checked that Eq.~(\ref{wzw}) reduces to $\mathcal{S}_\pi$ for $n_a = (\pi_1, \pi_2, \pi_3, \pi_4,1)$, and 
so $\mathcal{S}_{\mathrm{WZW}}$ is the explicitly SO(5) invariant form of $\mathcal{S}_\pi$. 

\section*{Theoretical consequences}

We now turn to a discussion of the theoretical consequences of the WZW term for the vicinity of the AF-VBS transition. For $q=1$, it has been
demonstrated in Refs.~\cite{levin,grover2,senthil3} that the O(5) nonlinear sigma model with O(3)$\times $O(2) anisotropy and a level 1
WZW term is equivalent to the CP$^1$ model in 2+1 dimensions.
This is the same model appearing in the AF-VBS transition of SU(2) quantum spin models \cite{senthil1,senthil2}, and is a 
relativistic field theory with a U(1) gauge field
and a two-component complex scalar $z_\alpha$. In terms of these fields, the AF order is $z_\alpha^\ast \sigma^s_{\alpha\beta} z_\beta$, with $s=x,y,z$;
so the vector AF order has been ``fractionalized'' into spinons $z_\alpha$. Alternatively, we can also view the $z_\alpha$ quanta 
as representing the vortices or anti-vortices in the VBS order \cite{levin} which, as we have just seen, carry spin $S=1/2$. 

Presently, the experimentally accessible case of the AF-VBS transition is in bilayer graphene, so we focus now on the $q=2$
case. With a level 2 
WZW term, the VBS vortices carry spin $S=1$, and therefore we need complex scalar fields with three components: we write these as $Z_s$,  
with $s=x,y,z$. The field theory of the $Z_s$ quanta is now the CP$^2$ model with anisotropic quartic terms; such a field theory was considered in Ref.~\cite{grover1} in a different context:
\beq
\mathcal{L}_{cp} = |(\partial_\mu - i A_\mu) Z_s|^2 + g |Z_s|^2 + u_1 (|Z_s|^2)^2 + u_2 (Z_s^2) (Z_t^{\ast 2}) .\nonumber
\eeq
Here $\mu$ is a spacetime index, 
$A_\mu$ is the emergent U(1) gauge field, $g$ is the coupling which tunes the AF to VBS transition, and $u_{1,2}$ are quartic couplings. 
In terms of the degrees of freedom in $\mathcal{L}_{cp}$, the three-component AF order parameter is now
$i \epsilon_{stu} Z_t^\ast Z_u$, while the complex VBS order $\langle \rho^x + i \rho^y \rangle \sim e^{i \theta}$ is 
the monopole operator in the U(1) gauge field \cite{vbsprb,senthil2}.

For both the CP$^1$ and CP$^2$ models mentioned above, both first and second order transitions are possible between
the AF and VBS states. A recent numerical study \cite{so5} on a single layer model indicates a first order transition for 
the parameters studied.

\section*{Experimental implications}

Finally, we turn to experimental consequences for bilayer graphene. The defining characteristic of deconfined criticality
is the presence of a gapless ``photon'' excitation of an emergent U(1) gauge field \cite{senthil1}. This is associated with
the $A_\mu$ above, and can also be interpreted as a ``spin-wave'' excitation
involving fluctuations of the angle $\theta$. Our definition of $\theta$ shows that it is the angular phase associated with off-diagonal-long-range
order (ODLRO) in valley space. The valley anisotropy terms in graphene are very small \cite{khari1,so5}, 
because it is suppressed by powers of the lattice spacing to the magnetic length; so we expect
a nearly gapless $\theta$ spin-wave mode to be present (and most of the remarks below to also apply) even in the case of a first-order transition.

Now recall the fact, noted earlier, that in the zero energy Landau levels the valley index coincides with the layer index
of bilayer graphene (and also the sublattice index). So ODLRO in valley space is accompanied by ODLRO in the layer space, i.e.,
$\theta$ is also the angular phase of interlayer excitonic superfluidity. 
Signatures of excitonic superfluidity have been observed in quantum Hall states in 
GaAs bilayers \cite{jim}. However, in the GaAs bilayers there is negligible tunneling of electrons between the layers, and this crucial
to the emergence of a U(1) symmetry which is broken by the excitonic condensate. So it might seem surprising that a
similar superfluidity can be present in graphene bilayers, in the presence of very strong tunneling between the layers. 
The resolution is the identification of the layer and valley indices in the zero energy Landau levels of bilayer graphene: 
in the absence of intervalley scattering by impurities,
and the irrelevancy of valley anisotropy terms to be presented below, 
there is also an emergent interlayer U(1) symmetry in bilayer graphene.

The counterflow electrical current can be written in terms of the gauge field
\beq
J_{t\mu} - J_{b\mu} = \frac{4e}{2 \pi} \epsilon_{\mu\nu\lambda} \partial_\nu A_\lambda \label{ct}
\eeq
where $J_t$ and $J_b$ are the currents in the top and bottom layers, and $e$ is the charge of the electron.
The factor of 4 is deduced from Eq.~(\ref{charge}), which shows that a AF Skyrmion in $(n_1, n_2, n_3)$ has excitonic charge density
$\langle \rho^z \rangle = 4$ for $q=2$.
The counterflow conductivity can be computed from Eq.~(\ref{ct}) using $\mathcal{L}_{cp}$; at the $g=g_c$ deconfined quantum critical
point this implies a universal value of order, the quantum unit of conductance $e^2 /h$.

For $g>g_c$, in the VBS state, the conductivity should be computed using an effective action for $\theta$ which includes the influence
of monopoles. Now the current is
\beq
J_{t\mu} - J_{b\mu} = 4 e \rho_s \partial_\mu \theta
\eeq
where $\rho_s$ is the stiffness of the excitonic superfluidity appearing the effective Lagrangian density
\beq
\mathcal{L}_\theta = \frac{\rho_s}{2} ( \partial_\mu \theta)^2 - y_3 \, \cos (3 \theta),
\eeq
with $y_3$ the fugacity of tripled monopoles which are allowed by the threefold rotational symmetry of the underlying honeycomb lattice \cite{vbsprb}.
The stiffness vanishes as $\rho_s \sim (g-g_c)^\nu$ by the Josephson relation, where $\nu$ is the correlation length exponent.
Away from the critical point, the bare value $y_3^0$ is proportional to the very small threefold valley anisotropy term \cite{khari1,so5}; 
$y_3$ has a further suppression \cite{CSY,senthil2} from the critical fluctuations of $\mathcal{L}_{cp}$ 
leading to $y_3 = y_3^0 (g-g_c)^{\nu \Delta}$, 
where $\Delta$ is the scaling dimension of the tripled monopole operator. So the effective ``interlayer tunneling'' term, $y_3$, is highly
suppressed near the deconfined quantum critical point. We also note that numerical studies on square lattice antiferromagnets have provided 
striking evidence for the emergent U(1) symmetry due to the suppression of monopoles \cite{sandvik}, and there is direct evidence for the 
suppression of monopoles on the honeycomb lattice in the work of Block {\em et al.\/} \cite{block}.

In GaAs bilayers \cite{jim}, the excitonic superfluidity is most directly observed in counterflow experiments, where
electric currents flow in opposite direction in the two layers. This would be technically more difficult in bilayer graphene, given the close spacing
of the layers, but experiments of this type would be ideal. 
In the bilayer graphene experiments of Weitz {\em et al.\/} \cite{weitz}, there is a Zeeman coupling to the magnetic field (whose consequences
have been studied earlier \cite{senthil2}), and an electric field is applied transverse to the layers. The electric field 
provides a small breaking of the layer-exchange symmetry. In the presence of such a symmetry breaking, there is a coupling
between the counterflow and parallel current modes, and a vestige of the counterflow superfluidity would also be present in a measurement of the
total current in both layers. 
Weitz {\em et al.} observe a phase transition out of the (presumed) AF state, signaled by the enhancement
of the conductivity. We propose that this enhancement is due to the coupling to counterflow superfluidity.\\

\noindent
{\bf Acknowledgements:} We thank D.~Abanin, Eun-Gook Moon, E.~Shimshoni, T.~Senthil, I.~Sodemann, and A.~Yacoby for useful 
discussions. This research
was supported by the NSF under Grant No. DMR-1360789, the Templeton foundation, and MURI Grant No. W911NF-14-1-0003 from ARO.
Research at Perimeter Institute is supported by the
Government of Canada through Industry Canada and by the Province of
Ontario through the Ministry of Research and Innovation.
J. L. is also supported by the STX Foundation.

\newpage 

\begin{appendix}

\begin{widetext}

\setcounter{equation}{0}
\setcounter{figure}{0}
\setcounter{table}{0}
\renewcommand{\theequation}{S\arabic{equation}}
\renewcommand{\thesection}{S\Roman{section}}
\renewcommand{\thefigure}{S\arabic{figure}}

\vspace{1mm}
\section*{Appendix A: Derivation of Skyrmion charge}

We give a detailed derivation of Eq.~(5). We work in Landau gauge and separate the quantum number $l$ in Eq.~(2) into $n$ and $X$, where $n$ is the Landau level index and $X$ is the center of the magnetic oscillator. For monolayer graphene ($q=1$) $n$ is 0, and for bilayer graphene ($q=2$) $n$ is over 0 and 1. There are $N_\Phi$ of $X$ labels, so the total number of states in the zero energy Landau level is $q N_\Phi$. 

The spatial wavefunction $\psi_{n, X} (\br)$ in Eq.~(2) is: 
	\begin{align}
		\psi_{n, X} (\br) =  e^{i X y} \varphi_{n, X} (x),
	\end{align}
where $\varphi_{n,X} (x)$ are the $n$-th harmonic oscillator eigenfunction:
	\begin{align}
		\varphi_{0,X} (x) &= \frac{1}{\pi^{1/4}} e^{-(x-X)^2/2}, \nn
		\varphi_{1,X} (x) &= \frac{\sqrt{2} (x-X)}{\pi^{1/4}} e^{-(x-X)^2/2}.
	\end{align}
Note that we work in the unit where the magnetic length $l_B = 1$. We also set the volume of the system as 1 to avoid volume factors in Fourier transform.

Now we can write the Hamiltonian in $c_{n, X}$ basis. 
	\begin{align}
		H &= -\lambda \int d^2 {\br} ~n_a (\br) \Psi^{\dagger} (\br)  \Gamma_a \Psi (\br) \nn
		&= -\lambda \int d^2 {\br}\, d^2 {\bq} ~  e^{- i \bq \cdot \br} n_a (- \bq) 
		\sum_{n,X,n',X'} c^{\dagger}_{n,X} \varphi_{n,X} (x) e^{-i X y}  \, \Gamma_a \,e^{i X' y} \varphi_{n',X'} (x) c_{n',X'}  \nn
		&= -\lambda \int d^2 {\bq} ~n_a (- \bq)  \sum_{n,n',X} e^{-i q_x X} c^{\dagger}_{n,X-q_y/2} \, \Gamma_a \, c_{n',X+q_y/2} F_{n,n'} (\bq) 
	\label{eq:H}
	\end{align}
The form factors $F_{n,n'}(\bq)$ are calculated as, 
\bea
F_{00} (\bq) &=& e^{-q^2/4} \nn
F_{01} (\bq) &=& \frac{(-i q_x - q_y)}{\sqrt{2}} e^{-q^2/4} \nn
F_{10} (\bq) &=& \frac{(-i q_x + q_y)}{\sqrt{2}} e^{-q^2/4} \nn
F_{11} (\bq) &=& (1-q^2/2) e^{-q^2/4} .
\eea
The difference in Eq.~(\ref{eq:H}) between $q=1$ and $q=2$ case comes from the summation limit of $n$ and $n'$.

Let us define the operator ${\mathcal O}_a (\bq)$ to be the last part of Eq.~(\ref{eq:H}). 
\begin{align}
	{\mathcal O}_a (\bq) = \sum_{n,n',X} e^{-i q_x X} c^{\dagger}_{n,X-q_y/2} \, \Gamma_a \, c_{n',X+q_y/2} F_{n,n'} (\bq) 
\end{align}
We can calculate the commutator of two ${\mathcal O}_a$ operators to the lowest order in momentum. In this section, we concentrate on the case of bilayer graphene, $q = 2$.
\begin{align}
	\left[{\mathcal O}_a (\bq) , {\mathcal O}_b (\bk) \right] 
	=& \sum_{n, n', m, Z} e^{-i(q_x + k_x)Z} e^{i(q_x k_y - q_y k_x)/2} F_{nm}(\bq) F_{mn'} (\bk) c^{\dagger}_{n, Z-(q_y + k_y)/2} \, \Gamma_a \Gamma_b \, c_{n', Z+(q_y +k_y)/2} - (\bq \leftrightarrow \bk , a \leftrightarrow b) \nn
	=& \sum_{Z} e^{-i(q_x + k_x)Z} c^{\dagger}_{1, Z-(q_y + k_y)/2} \left( \frac{\Gamma_a \Gamma_b + \Gamma_b \Gamma_a}{2} \right) c_{1, Z+(q_y +k_y)/2} \times \left( 2 i (k_y q_x - k_x q_y ) + {\mathcal O}(\bq, \bk)^2 \right) \nn
	&+  \sum_{n, Z} e^{-i(q_x + k_x)Z} c^{\dagger}_{n, Z-(q_y + k_y)/2} \left( \frac{\Gamma_a \Gamma_b - \Gamma_b \Gamma_a}{2} \right) c_{n, Z+(q_y +k_y)/2} \times \left( 2  + {\mathcal O}(\bq, \bk) \right) 
	\label{eq:comm}
\end{align}

For the 3-vector field $\vec{N} (\br)$ defined in the main text, we assume $n_w \rightarrow 1$ as $\br \rightarrow \infty$, i.e., $\vec{N} (\br \rightarrow \infty) = (0, 0, 1) \equiv \vec{N}_0 $. This choice is arbitrary and may be rotated by any SO(3) transformation. If we denote $|GS\rangle$ to be the ground state with $\vec{N} (\br) = \vec{N}_0$, the expectation value of $\Psi^{\dagger}(\br) \Gamma_w \Psi(\br)$ will be, 
\begin{align}
	\langle GS| \Psi^{\dagger}(\br) \Gamma_w \Psi(\br) |GS \rangle = \frac{4}{2\pi}.
	\label{eq:w_expect}
\end{align}
We impose a Skyrmion texture to the ground state by rotating the spinors by an operator ${\mathcal O}$,
\begin{align}
	{\mathcal O} &= \sum_{\br} \vec{\Omega} (\br) \cdot \left( \Psi^{\dagger}(\br) \frac{\vec{\Gamma}}{2} \Psi(\br) \right) \nn
	&= \sum_{\bq} \frac{\vec{\Omega} (-\bq)}{2} \vec{\mathcal O}_N (\bq),
\end{align}
where $\vec{\Omega} (\br) = \vec{N}_0 \times \vec{N} (\br) $, $\vec{\Gamma} = (\Gamma_u , \Gamma_v , \Gamma_w)$, and $\vec{\mathcal O}_N = ({\mathcal O}_u , {\mathcal O}_v , {\mathcal O}_w )$. The Skyrmion state is now $e^{-i {\mathcal O}}| GS\rangle$. Let us also define the Fourier transform of the operator $\Psi^{\dagger}(\br) i \Gamma_u \Gamma_v \Gamma_w \Psi(\br)$ to be ${\mathcal O}_\Gamma (\bk)$.

Now we follow the derivation from Ref. [39]. 
\begin{align}
	\langle {\mathcal O}_\Gamma (\bk) \rangle 
	&= \langle GS| e^{i \mathcal O} {\mathcal O}_\Gamma (\bk) e^{-i \mathcal O} |GS \rangle - \langle GS| {\mathcal O}_\Gamma (\bk) |GS \rangle \nn
	&= i \langle GS| [{\mathcal O}, {\mathcal O}_\Gamma (\bk)] | GS \rangle - \frac{1}{2} \langle GS| \left[{\mathcal O}, [{\mathcal O}, {\mathcal O}_\Gamma (\bk)]\right]| GS \rangle + \cdots
\end{align}
The first term vanishes and we ignore the higher order terms in ${\mathcal O}$ since we consider a smooth texture. The derivation boils down to calculating the ground state expectation value of $\left[{\mathcal O}, [{\mathcal O}, {\mathcal O}_\Gamma (\bk)]\right]$. 

First, we compute $ [{\mathcal O}, {\mathcal O}_\Gamma (\bk)]$ using Eq.~(\ref{eq:comm}). 
\begin{align}
	[{\mathcal O}, {\mathcal O}_\Gamma (\bk)] 
	&= \sum_\bq \frac{\Omega_a (-\bq)}{2}[{\mathcal O}^a_N (\bq), {\mathcal O}_\Gamma (\bk)] \nn
	&= \sum_\bq \frac{\Omega_a (-\bq)}{2} \left( - k_y q_x + k_x q_y \right) \epsilon_{acd} \sum_{Z} e^{-i(q_x + k_x)Z} c^{\dagger}_{1, Z-(q_y + k_y)/2} \, \Gamma_c \Gamma_d \, c_{1, Z+(q_y +k_y)/2}
\end{align}
$a = u, v, w$ is the vector index and repeated indices are summed implicitly.
Since $[\Gamma_a , \Gamma_u \Gamma_v \Gamma_w] = 0$ for all $a$, we considered only the symmetric term in Eq.~(\ref{eq:comm}). The anti-symmetric tensor is ordered as $\epsilon_{uvw} = 1$. 

Now we calculate $\left[{\mathcal O}, [{\mathcal O}, {\mathcal O}_\Gamma (\bk)]\right]$. 
\begin{align}
	\left[{\mathcal O}, [{\mathcal O}, {\mathcal O}_\Gamma (\bk)]\right]
	&= \sum_\bp \frac{\Omega_b (-\bp)}{2} \left[{\mathcal O}^b_N (\bp), [{\mathcal O}, {\mathcal O}_\Gamma (\bk)]\right] \nn
	&= \sum_{\bp , \bq} \frac{\Omega_a (-\bq) \Omega_b (-\bp)}{4} \left( - k_y q_x + k_x q_y \right) \epsilon_{acd} \left[ \sum_{n,n',X} e^{-i p_x X} c^{\dagger}_{n,X-p_y/2} \, \Gamma_b \, c_{n',X+p_y/2} F_{n,n'} (\bp) , \right. \nn
	&\qquad \qquad \qquad \qquad \qquad \qquad \qquad \qquad \qquad \qquad
	\left. \sum_{Z} e^{-i(q_x + k_x)Z} c^{\dagger}_{1, Z-(q_y + k_y)/2} \,  \Gamma_c \Gamma_d \,c_{1, Z+(q_y +k_y)/2} \right]
\end{align}
We have assumed the ground state is uniform, $\vec{N} (\br) = \vec{N}_0$. Therefore, the expectation value of the above expression is nonzero only when $- \bp = \bq + \bk$. When this condition is satisfied, the last commutator simplifies to $\sum_{Z} c^{\dagger}_{1, Z}[\Gamma_b , \Gamma_c \Gamma_d] c_{1, Z}$, to the lowest order of momentum. Also using $\epsilon_{acd} [\Gamma_b , \Gamma_c \Gamma_d] = 4 \epsilon_{abc} \Gamma_c$, we obtain the expression for $\langle {\mathcal O}_\Gamma (\bk) \rangle$.
\begin{align}
	\langle {\mathcal O}_\Gamma (\bk) \rangle
	&= \sum_{\bq}\epsilon_{abc}  \left( k_y q_x - k_x q_y \right) \frac{\Omega_a (-\bq) \Omega_b (\bq + \bk)}{2} \sum_Z \langle GS | c^{\dagger}_{1, Z} \,\Gamma_c \, c_{1, Z} | GS \rangle \nn
	&= \sum_{\bq}\epsilon_{ab}  \left( k_y q_x - k_x q_y \right) \frac{\Omega_a (-\bq) \Omega_b (\bq + \bk)}{2} \sum_Z \langle GS | c^{\dagger}_{1, Z} \,\Gamma_w \, c_{1, Z} | GS \rangle \nn
	&= \frac{1}{2\pi}\sum_{\bq} \epsilon_{ab} \left( k_y q_x - k_x q_y \right) \Omega_a (-\bq) \Omega_b (\bq + \bk) \nn
	&= \frac{1}{2\pi}\sum_{\bq} \epsilon_{ab} \left( -i \bq \Omega_a (-\bq) \right) \times \left(i(\bk + \bq) \Omega_b (\bq + \bk) \right) \cdot \hat{z}
\end{align}
The second equality make use of the expectation value being nonzero only when $\Gamma_c = \Gamma_w$ in the defined ground state. The new anti-symmetric tensor has indicies $a, b = u, v$, and $\epsilon_{uv} = 1$. Note that the expectation value is half of Eq.~(\ref{eq:w_expect}) since only $n=1$ contributes in the above equation. We obtain the final result by Fourier transforming back to real space. 
\begin{align}
	\langle \Psi^{\dagger}(\br) i \Gamma_u \Gamma_v \Gamma_w \Psi(\br) \rangle 
	&= \frac{1}{2\pi}  \epsilon_{ab}  \left( \nabla \Omega_a (\br) \times \nabla \Omega_b (\br) \right) \cdot \hat{z} \nn
	&= \frac{2}{2\pi} \vec{N} \cdot \left( \partial_x \vec{N} \times \partial_y \vec{N} \right)
\end{align}
This is Eq.~(5) for $q=2$. The simpler $q=1$ case follows from the exactly same procedure. 

\section*{Appendix B: Diagrammatic derivation of WZW term}

We turn to a detailed derivation of Eq. (7). As in the main text, consider the situation $n_a (\br)= (\pi_1, \pi_2, \pi_3, \pi_4, 1)$ where $n_a$ is polarized near $(0,0,0,0,1)$. Substituting this to Eq.~(\ref{eq:H}) we get the Hamiltonian, 
	\begin{align}
		H &= H_0 + H_\pi \nn
		&= - \lambda \sum_{n, X} c^{\dagger}_{n, X} \Gamma_5 c_{n, X} 
		- \lambda \sum_{i = 1}^4 \int d^2 \bq \, \pi_i (-\bq) \sum_{n, n', X} e^{-i q_x X} c^{\dagger}_{n, X-q_y/2} \Gamma_i \,c_{n', X+q_y/2} F_{n, n'}(\bq) .
		\label{eq:Hp}
	\end{align}

From the above equation, we integrate out the fermions to get an effective theory for the fluctuating order parameters. The coupling between the order parameters appear at fourth order of one-loop expansion. In momentum space, the four point coupling between the order parameter fields are, 
\begin{align}
	S_1 =  \int \prod_{\alpha=1}^{3} d {\bp}_\alpha K^{j k m; i}_{{\bp}_1 {\bp}_2 {\bp}_3}
		\pi_{j} ({\bp}_1 ) \pi_{k}({\bp}_2) \pi_{m}({\bp}_3)\pi_{i}(-{\bp}_1 -{\bp}_2 -{\bp}_3  ).
	\label{eq:s1}
\end{align}
Among these terms, we are most interested in the topological term,
\begin{align}
	S_{\pi} = i K \int d^2 \br d\tau \epsilon_{i j k m} \pi_i \partial_x \pi_j \partial_y \pi_k \partial_\tau \pi_m.
	\label{eq:topo_action}
\end{align}

To extract the coefficient $K$ from $S_1$ we expand $K^{j k m; i}_{{\bp}_1 {\bp}_2 {\bp}_3}$ in powers of momenta and frequency, and consider the terms linear in $p_1 p_2 p_3$:
\begin{align}
	K^{j k m; i}_{{\bp}_1 {\bp}_2 {\bp}_3} = \cdots + K^{j k m; i}_{\alpha \beta \gamma} p_1^\alpha p_2^\beta p_3^\gamma+ \cdots.
\end{align}
Here, $\alpha, \, \beta, \, \gamma$ are spacetime indices $\tau, \, x, \, y$. In real space, these terms correspond to the derivative expansion. 
\begin{align}
	S_1 = \cdots + i \sum_{i, j, k, m = 1}^4 K^{j k m; i}_{\alpha \beta \gamma} \int  d^2 \br d\tau  
		\pi_{i} \partial_{\alpha} \pi_{j} \partial_{\beta} \pi_{k} \partial_{\gamma} \pi_{m}+ \cdots
	\label{eq:action_expand}
\end{align}
Comparing Eq.~(\ref{eq:topo_action}) with Eq.~(\ref{eq:action_expand}), we obtain the expression for $K$ in terms of $K^{j k m; i}_{\alpha \beta \gamma}$'s. 
\begin{align}
	24 K &= \epsilon_{\alpha \beta \gamma} \epsilon_{i j k m} K^{j k m ; i}_{\alpha \beta \gamma}
	\label{eq:k}
\end{align}
Note the summation of repeated indices are implicit, and thus the right-hand side of the above equation consists of 144 terms. 

Now we only need to calculate $K^{j k m; i}_{\alpha \beta \gamma}$ to obtain $K$. This can be done by calculating the 
box diagrams in Fig.~2. The kinetic energy of fermions are quenched, and the propagators are momentum independent. Therefore, the momentum dependence comes from the vertices and frequency dependence comes from the propagator. 

Recalling Eq.~(\ref{eq:Hp}), the inverse Green's function is written as $G^{-1} = -i \omega +H_0$, where $\omega$ is the Matsubara frequency of the fermions. The effective action of order $\pi^4$ in perturbation theory, which is of our interest according to Eq.~(\ref{eq:action_expand}), is:
\begin{align}
	\mathcal{S} [\pi^4]= \frac{1}{4} \textrm{tr}  \left( G H_\pi G H_\pi G H_\pi G H_\pi \right).
\end{align}
Considering the momentum flow in Fig.~2, we obtain the expression for $K^{j k m; i}_{\alpha \beta \gamma}$.
\begin{align}
	 K^{j k m; i}_{\alpha \beta \gamma} &= \frac{1}{4} \left( \partial_{p_1^\alpha} \partial_{p_2^\beta} \partial_{p_3^\gamma}\textrm{tr} \left.  \left(  G(k) H_\pi^j(p_1) G(k+p_1) H_\pi^k(p_2) G(k+p_1 +p_2) \right. \right. \right. \nn
	 &\phantom{~~~~~~~~~~~~~~~~~~~~~~~~~} \left. \left. \left. \times H_\pi^m(p_3) G(k+ p_1 + p_2 +p_3) H_\pi^i(-p_1 -p_2 -p_3) \right)\right) \right|_{ p_1 =  p_2 = p_3 = 0 }
	 \label{eq:top_exp}
\end{align}
We used a shorthand notation $H_\pi^a$ for $H_\pi^a = H_\pi (\pi_{i = a} = 1, \pi_{i \neq a} = 0)$. 

We get the value of $K$ by calculating Eq.~(\ref{eq:top_exp}). In zero temperature, the frequency integral can be done analytically,
	\begin{align}
		K &= \frac{1}{2\pi} \int d \omega \sum_{X} \frac{2q \lambda^5}{(\omega^2 + \lambda^2)^3} \nn
		&= \sum_X \frac{3q}{8} \nn
		&= \frac{3q}{16\pi}.
	\end{align}
The individual diagram is independent of $X$, merely reflecting momentum conservation. Therefore the $X$ summation is just multiplying $N_\Phi$, 
which equals (sample area)/$({2\pi})$. This gives the result in Eq.~(7). 
\\

\end{widetext}

\end{appendix}


\begin{thebibliography}{99}

\bibitem{weitz} R. T. Weitz, M. T. Allen, B. E. Feldman, J. Martin, and A. Yacoby,
Science {\bf 330}, 812 (2010).

\bibitem{freitag} F. Freitag, J. Trbovic, M. Weiss, and C. Sch\"onenberger, Phys. Rev. Lett. {\bf 108}, 076602 (2012). 

\bibitem{macdonaldexp} J.~Velasco, L.~Jing, W.~Bao,	 Y.~Lee, P.~Kratz, V.~Aji, M.~Bockrath,
C.~N.~Lau, C.~Varma, R.~Stillwell, D.~Smirnov, F. Zhang, J.~Jung, and A.~H.~MacDonald, Nat.
Nanotech. {\bf 7}, 156 (2012).

\bibitem{young1} A.~F.~Young, C. R. Dean, L. Wang, H. Ren, P. Cadden-Zimansky, K. Watanabe, T. Taniguchi, 
J. Hone, K.~L.~Shepard and P. Kim, Nat. Phys. {\bf 8}, 550 (2012).

\bibitem{maher} P.~Maher, C. R. Dean,  A.~F.~Young, T. Taniguchi, K. Watanabe,
K. L. Shepard, J.~Hone and P. Kim, Nat. Phys. {\bf 9}, 154 (2013).

\bibitem{basel} F.~Freitag, M.~Weiss, R.~Maurand, J.~Trbovic, and C.~Sch\"onenberger
Phys. Rev. B {\bf 87}, 161402(R) (2013).

\bibitem{young2} A. F. Young, J. D. Sanchez-Yamagishi, B. Hunt, S. H. Choi, K. Watanabe, T. Taniguchi, R. C. Ashoori, 
and P.~Jarillo-Herrero, Nature (London) {\bf 505}, 528 (2014).

\bibitem{herbut1} I.~F.~~Herbut, Phys. Rev. B {\bf 76}, 085432 (2007).

\bibitem{jung} J.~Jung and A.~H.~MacDonald, Phys. Rev. B {\bf 80}, 235417 (2009).

\bibitem{vafek} V.~Cvetkovic, R.~E.~Throckmorton, and O.~Vafek, Phys. Rev. B {\bf 86}, 075467 (2012).

\bibitem{allan} Fan Zhang, Hongki Min, and A.~H.~MacDonald, Phys. Rev. B {\bf 86}, 155128 (2012).

\bibitem{levitov} R.~Nandkishore and L.~Levitov, Phys. Rev. B {\bf 82}, 115431 (2010).

\bibitem{khari1} M.~Kharitonov, Phys. Rev. B {\bf 85}, 155439 (2012).

\bibitem{khariprl} M.~Kharitonov, Phys. Rev. Lett. {\bf 109}, 046803 (2012).

\bibitem{khari2} M.~Kharitonov, Phys. Rev. B {\bf 86}, 075450 (2012).

\bibitem{khari3} M.~Kharitonov, Phys. Rev. B {\bf 86}, 195435 (2012).

\bibitem{so5} Fengcheng Wu, I.~Sodemann, Y.~Araki, A.~H.~MacDonald, and 
T.~Jolicoeur, Phys. Rev. B {\bf 90}, 235432 (2014).

\bibitem{JLSS14} Junhyun Lee and S.~Sachdev, Phys. Rev. B {\bf 90}, 195427 (2014). 

\bibitem{efrat} K.~Dhochak, E.~Shimshoni, and E.~Berg, Phys. Rev. B {\bf 91}, 165107 (2015).

\bibitem{vbsprb} N. Read and S. Sachdev, Phys. Rev. B {\bf 42}, 4568 (1990).

\bibitem{senthil1}  T.~Senthil, A.~Vishwanath, L.~Balents, S.~Sachdev, and
M.~P.~A.~Fisher, Science {\bf 303}, 1490 (2004).

\bibitem{senthil2} T.~Senthil, L.~Balents, S.~Sachdev, A.~Vishwanath,
and M.~P.~A.~Fisher, Phys. Rev. B  {\bf 70}, 144407 (2004).

\bibitem{clark} B.~K.~Clark, D.~A.~Abanin and S.~L.~Sondhi, Phys. Rev. Lett.
{\bf 107}, 087204 (2011).

\bibitem{lauchli} A.~F.~Albuquerque, D.~Schwandt, B.~Het\'enyi, S.~Capponi, M.~Mambrini, and A.~M.~L\"auchli, 
Phys. Rev. B {\bf 84}, 024406 (2011).

\bibitem{ganesh} R.~Ganesh, J.~van den Brink, and S. Nishimoto,
Phys. Rev. Lett. {\bf 110}, 127203 (2013).

\bibitem{zhu} Z. Zhu, D.~A.~Huse, and S.~R.~White,
Phys. Rev. Lett. {\bf 110}, 127205 (2013).

\bibitem{block} M.~S.~Block, R.~G.~Melko, and R.~K.~Kaul, Phys. Rev. Lett. {\bf 111}, 137202 (2013).

\bibitem{shengfisher} S.-S. Gong, D.~N.~Sheng, O.~I.~Motrunich, and M.~P.~A.~Fisher, Phys. Rev. B {\bf 88}, 165138 (2013).

\bibitem{damle} S.~Pujari, K.~Damle, and F.~Alet, Phys. Rev. Lett. {\bf 111}, 087203 (2013).

\bibitem{langsun}  T.~C.~Lang, Z. Y. Meng, A.~Muramatsu, S.~Wessel, and 
F.~F.~Assaad, Phys. Rev. Lett. {\bf 111}, 066401 (2013).

\bibitem{WZ} J.~Wess and B.~Zumino, Phys. Lett. B {\bf 37}, 95 (1971)

\bibitem{Witten} E.~Witten, Nucl. Phys. {\bf B223}, 422 (1983).

\bibitem{abanov} A. G. Abanov and P. B. Wiegmann,
Nucl. Phys. {\bf B570}, 685 (2000).

\bibitem{tanakahu} A.~Tanaka and X.~Hu, Phys. Rev. Lett. {\bf 95}, 036402 (2005).

\bibitem{senthil3} T. Senthil and M.~P.~A.~Fisher,
Phys. Rev. B {\bf 74}, 064405 (2006).

\bibitem{grover2} T.~Grover and T.~Senthil, Phys. Rev. Lett. {\bf 100}, 156804 (2008).

\bibitem{liang} L. Fu, S.~Sachdev, and C. Xu, 
Phys. Rev. B {\bf 83}, 165123 (2011). 

\bibitem{egmoon} Eun-Gook Moon, Phys. Rev. B {\bf 85}, 245123 (2012).

\bibitem{moon} K.~Moon {\em et al.\/}, Phys. Rev. B {\bf 51}, 5138 (1995).

\bibitem{abanin}  D. A. Abanin, S. A. Parameswaran, and S. L. Sondhi,
Phys. Rev. Lett. {\bf 103}, 076802 (2009).

\bibitem{ref} See Supplemental Material at http://link.aps.org/\\supplemental/10.1103/PhysRevLett.114.226801 for detailed derivations of Eqs.~(5) and (7).

\bibitem{herbut2} I.~F.~Herbut, C.-K.~Lu, and B.~Roy, Phys. Rev. B {\bf 86}, 075101 (2012).

\bibitem{levin} M.~Levin and T. Senthil,
Phys. Rev. B {\bf 70}, 220403(R) (2004).

\bibitem{hou-prb} C.-Y. Hou, C.~Chamon, and C.~Mudry, 
Phys. Rev. B {\bf 81}, 075427 (2010). 

\bibitem{grover1} T.~Grover and T.~Senthil, Phys. Rev. Lett. {\bf 98}, 247202 (2007).

\bibitem{jim} J.~P.~Eisenstein, Annu. Rev. Condens. Matter Phys. {\bf 5}, 159 (2014).

\bibitem{CSY} A. V. Chubukov, S. Sachdev, and J. Ye, Phys. Rev. B {\bf 49}, 11919 (1994).

\bibitem{sandvik} A.~W.~Sandvik, Phys. Rev. Lett. {\bf 98}, 227202 (2007).



\end{thebibliography}
\end{document}